\documentclass[aps,prc,twocolumn,showpacs]{revtex4-1}
\usepackage{graphicx}
\begin{document}
\title{Baryon production and net-proton distributions in relativistic heavy ion collisions}
\author{C. B. Yang  and Xin Wang }
\affiliation{Institute of Particle Physics, Central China Normal
University,Wuhan 430079, People's Republic of China}
\affiliation{Key Laboratory of Quark and Lepton Physics (CCNU),
Ministry of Education, People's Republic of China}
\date{\today}
\begin{abstract}

  The higher order moments of the net-baryon distributions in relativistic
  heavy ion collisions are useful probes for the QCD critical point and fluctuations.
   We study the net-proton distributions and their moments in a simple model
  which considers the baryon stopping and pair production effects in the processes.
  It is shown that a single emission source model can explain the experimental data well.
  Centrality and energy dependence of the distributions
  and higher moments is discussed.

\pacs{25.75.Gz, 21.65.Qr}
\end{abstract}
\maketitle

\section{Introduction}
The investigation of QCD phase diagram is fundamental to our understanding of strong
interactions. At vanishing baryon chemical potential, lattice QCD calculations predict
the occurrence of a cross-over from hadronic phase to the quark-gluon plasma phase above
a critical temperature of about 170-190 MeV \cite{YA,JB}. A distinct singular feature of the phase
diagram is the QCD critical point \cite{MAS} which is located at the end of the transition boundary.
A characteristic feature of the critical point is the divergence of the correlation length $\xi$
and extremely large critical fluctuations. In ultra-relativistic heavy ion collisions, however,
because of finite size and rapid expansion of the system, those divergence may be washed out. As estimated
in \cite{MAS}, the critical correlation length in heavy ion collisions is not divergent but about
2-3 fm. Remnants of those critical large fluctuations may become accessible in heavy ion collisions
through an event-by-event analysis of fluctuations in various channels of conservative  hadron quantum
numbers, for example, baryon number, electric charge, and strangeness \cite{FLUC}.
In an energy scan there would be a non-monotonic behavior of non-Gaussian multiplicity fluctuations,
which would be a clear signature for the existence of a critical point. In fact, at vanishing chemical
potential it has been shown that moments of conservative charge fluctuations are sensitive indicators
for the occurrence of a transition from hadronic to partonic matter \cite{SE}.

  Recently, the higher order moments of net-baryon distributions in heavy ion collisions at RHIC
  energies have aroused great interest both experimentally \cite{STAR} and theoretically \cite{THEO1,THEO2,THEO3,THEO4}. Experimentally, neutrons can not be detected easily and
  the reconstruction efficiency is very low for strange hadrons. Fortunately, theoretical
  calculations confirmed that the net-proton distribution can be a meaningful observable for the
  purpose of detecting the critical fluctuations of net baryons in heavy ion collisions \cite{NP}.
  The theoretical interest on these higher order moments comes from the discovery of the relation
between the moments and the thermal fluctuations near the critical points. It is shown that
the higher order moments have stronger dependence on the correlation length $\xi$ and are therefore
more sensitive to the critical fluctuations. If some memory of large correlation length persists in
the thermal medium in hadronization process, this must be reflected in higher order moments of the
distributions. It has been predicted \cite{MAS2} that the third moment, called skewness, is  proportional
to $\xi^{4.5}$ and fourth moment, or kurtosis, proportional to $\xi^7$ while
  the second moment proportional to $\xi^2$. More importantly, the moments are closely related
  to the susceptibilities of the thermal medium. It has been argued that information of QCD
  phase diagram  and the critical point
  can be obtained from the energy dependence of those moments \cite{THEO1}.
  The moments of net-proton distributions are studied with different theoretical models such as
  AMPT and UrQMD \cite{THEO3}, HIJING \cite{LUO}, and hadron resonance gas model \cite{FK} etc.
  All those theoretical models are quite complicated and many microscopic processes are
  involved, and as a result many parameters can be tuned in the investigations. Therefore, the underlying
  physics behind the experimental results on the higher order moments of the net-proton
  distribution is not very transparent from the model studies. In addition, those studies focused
  on the moments only and made no direct comparison with the experimentally obtained distributions.

     In this paper, we will investigate the net-proton distributions in Au+Au collisions
  at $\sqrt{s_{NN}}=200 {\rm GeV}$
  from very simple physics considerations: baryon stopping and baryon pair production.
  These physics effects are well known from studying heavy ion collisions in the past decades.
  We will show that such simple physics can be used to reproduce the experimentally
  observed net-proton distributions at different colliding centralities with parameters chosen
  properly. Then higher moments can be calculated  numerically from the distributions.
  In this way the centrality dependence of those moments can be predicted.

  This paper is organized as follows. In next section, we will address the physics points in our
  considerations for an emission source. Analytical expressions for the net-proton distribution will
  be given.  The model will be used to fit the experimental data on the
  distribution.  The  model can fit the data nearly perfectly. The centrality
  dependence of the moments will then be predicted. Also the moments at LHC energies are discussed.
  The last section will be for a brief summary.

\section{Model consideration for an emission source}
It was well established that the net-baryon number would be zero in heavy ion collisions
in central rapidity region if there were no nuclear stopping in the processes.
Because the nuclear stopping effect depends on the collision energy and the size of the system,
 the baryon number stopped in a rapidity region is closely related to the number of
participant nucleons $N_{\rm part}$. In more central collisions the net baryon number
will be larger.  We  consider a case in which all final state baryons are assumed being
produced from one emission source. The initial mean nucleon number in the source is denoted
as $B$ which may be different for different colliding centralities.  Considering the randomness
and independence of the nucleon-nucleon collisions in a heavy ion collision, the probability of finding
$N_0$  baryons stopped in the kinematic region under investigation can be assumed, with the given
mean number $B$, as
\begin{equation}
P_0(N_0,B)=\frac{B^{N_0}}{N_0!}\exp(-B)\ .
\label{eq2}
\end{equation}
Out of those $N_0$ stopped nucleons, some of them are proton, others neutrons.
 The probability of finding
$N_p$ protons from $N_0$ nucleons is
\begin{equation}
Q_0(N_0,N_p)=C^{N_p}_{N_0} \rho^{N_p} (1-\rho)^{N_0-N_p}\
\label{eq3}
\end{equation}
with $\rho=Z/A$ the fraction of proton in the nucleus. The above formulas determine the distribution of
net proton number in the source in the initial state of the collisions.

Then one can consider the baryon production in the collisions. Baryons can be produced from
various channels. The baryon number is a conserved
observable, thus baryons must be produced in baryon-antibaryon pairs. The pair production
may be independent, and as a result of the independence, the probability for producing
 $M$ baryon pairs must be a Poissonian with the given mean number of produced pairs $\mu$ as
 a parameter
 \begin{equation}
P_0(M,\mu)=\frac{\mu^M}{M!}\exp(-\mu)\ .
\label{eq4}
\end{equation}
The parameter $\mu$  depends on the colliding centrality. For more central collisions, the
colliding system is larger, therefore $\mu$ should be
 larger, and more nucleon-antinucleon pairs can be produced in the process.

It is the right place to compare the number distributions used in this paper and in others. We use a
Poisson distribution for the baryon pair distribution, supposing the independent production of the pairs.
In \cite{chen} the distributions for both proton and anti-proton are assumed Poissonian, implying that protons
 and anti-protons are produced completely independently. Therefore the baryon number conservation may be
 violated in any event. In \cite{begun}, a canonical ensemble is employed to derive the number distribution
 for $\pi$ systems. This is reasonable because there are a lot of $\pi$ particles in the final state of
 heavy ion collisions. But a simple transportation of the method to the case for baryon production may
 be problematic, because the relevant baryon particle number may be not large enough for an
 equilibrium statistical description.

 In the strong production of nucleon-antinucleon pairs, isospin is conserved. Suppose that $N_1$ protons,
 $N_2$ anti-protons, $N_3$ neutrons and $N_4$ anti-neutrons are produced in the process, the conservation
 of isospin reads $N_1-N_2=N_3-N_4$ if the effect from the presence of mesons is neglected.
 Thus we have $N_1+N_4=N_2+N_3=M$.  The probability of finding $N_1$ protons can be assumed as
 \begin{equation}
 Q_1(N_1,M)=2^{-M}C_{M}^{N_1}\ .
\end{equation}
In writing this equality, we assume that all the produced
 pairs are within the kinematic range detected experimentally. Of course, this is a rather rough
 approximation. In fact, some of the produced nucleons can go out of that range and cannot be
 included in measuring the net-protons in the event. The effect from limited kinematical acceptance
 can be taken into account by introducing one more parameter for the probability of the
 produced baryon in the detected region. To avoid this complexity, in this paper,
 such an effect is effectively treated as having the number of pairs $M$ a little smaller in the
 event. Therefore, the value of the parameter $\mu$ obtained from the fitting in this paper should be a little
 bit smaller than the real one.
In the same way, the probability of finding $N_2$ anti-proton is $Q_1(N_2,M)$.
Then the distribution of net-proton $\Delta p$ from an emission source can be expressed as
\begin{eqnarray}
P(\Delta p) =\sum_{N_0,N_p,M, N_1,N_2} P_0(N_0,B)Q_0(N_0,N_p)\nonumber\\
 Q_1(N_1, M) Q_1(N_2,M)P_0(M,\mu) \delta_{\Delta p, N_p+N_1-N_2}\ .
\end{eqnarray}
By inserting an identity expression $\delta_{m,n}=\int_0^{2\pi}dxe^{i(m-n)x}/2\pi$, the above equation
can be rewritten as
\begin{eqnarray}
P(\Delta p)&=&\int_0^\pi \frac{dx}{\pi} e^{-(2B\rho+\mu)\sin^2\frac{x}{2}}\cos(x\Delta p-B\rho\sin x)\ .\label{eq_sin}
\end{eqnarray}
As can be seen from the above expression, the net-baryon distribution
 depends on two combined parameters, $B\rho$ and $\mu$ instead of $B, \rho$ and $\mu$ separately.
 One can check easily that $B\rho$ is the mean value of the distribution $P(\Delta p)$.

\section{Comparison with the experimental data}
\begin{figure}[tbph]
\includegraphics[width=0.45\textwidth]{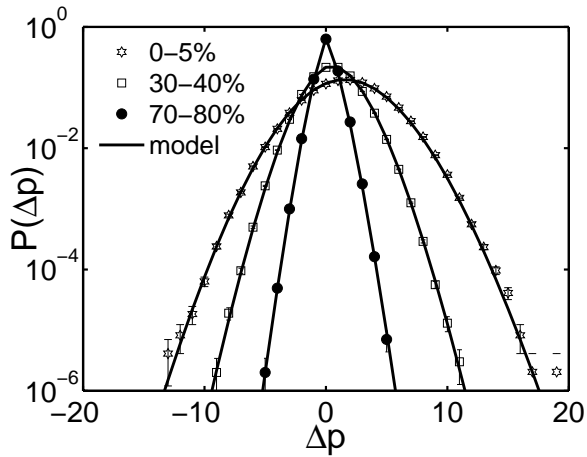}
\caption{The net-proton distributions from single emission source. The points are from Ref. \cite{STAR},
and the curves are calculated from Eq. \ref{eq_sin}.}
\label{fig1}
\end{figure}

The expression Eq. (\ref{eq_sin}) enables us to compare the calculated net-proton distributions from an
emission source to the experimental data from STAR \cite{STAR}, as shown in Fig. \ref{fig1}.
The parameters used are tabulated in TABLE \ref{tab1}. The parameters show the expected
behaviors from central to peripheral collisions. As one can see from the figure, the agreement with
the data is very good over five orders of magnitude.
 Almost all calculated points for $F(\Delta p)$ lie within the experimental error bars.

\begin{table}
  \centering
  \begin{tabular}{|c|c|c|c|}\hline\hline
  centrality & $N_{\rm part}$ & $\mu$ & $B\rho$ \\ \hline
  0-5\% & 351.4 & 14.5 & 1.65 \\ \hline
  30-40\% & 114.2 & 5.5 & 0.632 \\ \hline
  70-80\% & 13.4 & 0.83 & 0.075 \\ \hline\hline
  \end{tabular}
  \caption{Fitted parameters for Fig. \ref{fig1}}\label{tab1}
\end{table}

\begin{figure}[tbph]
\includegraphics[width=0.45\textwidth]{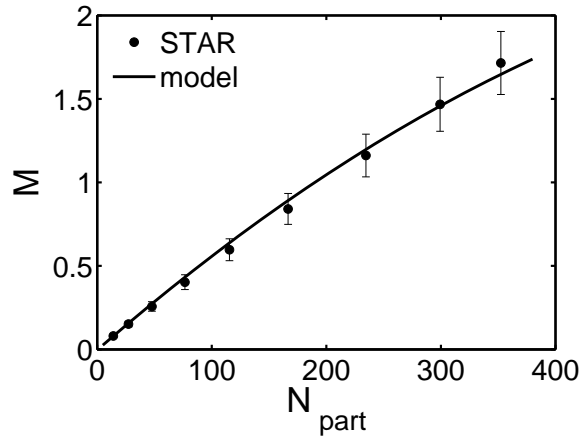}
\caption{The mean net-proton from multiple emission sources. The points are from Ref. \cite{STAR},
and the curve is from our model calculation.}
\label{fig3}
\end{figure}

\begin{figure}[tbph]
\includegraphics[width=0.45\textwidth]{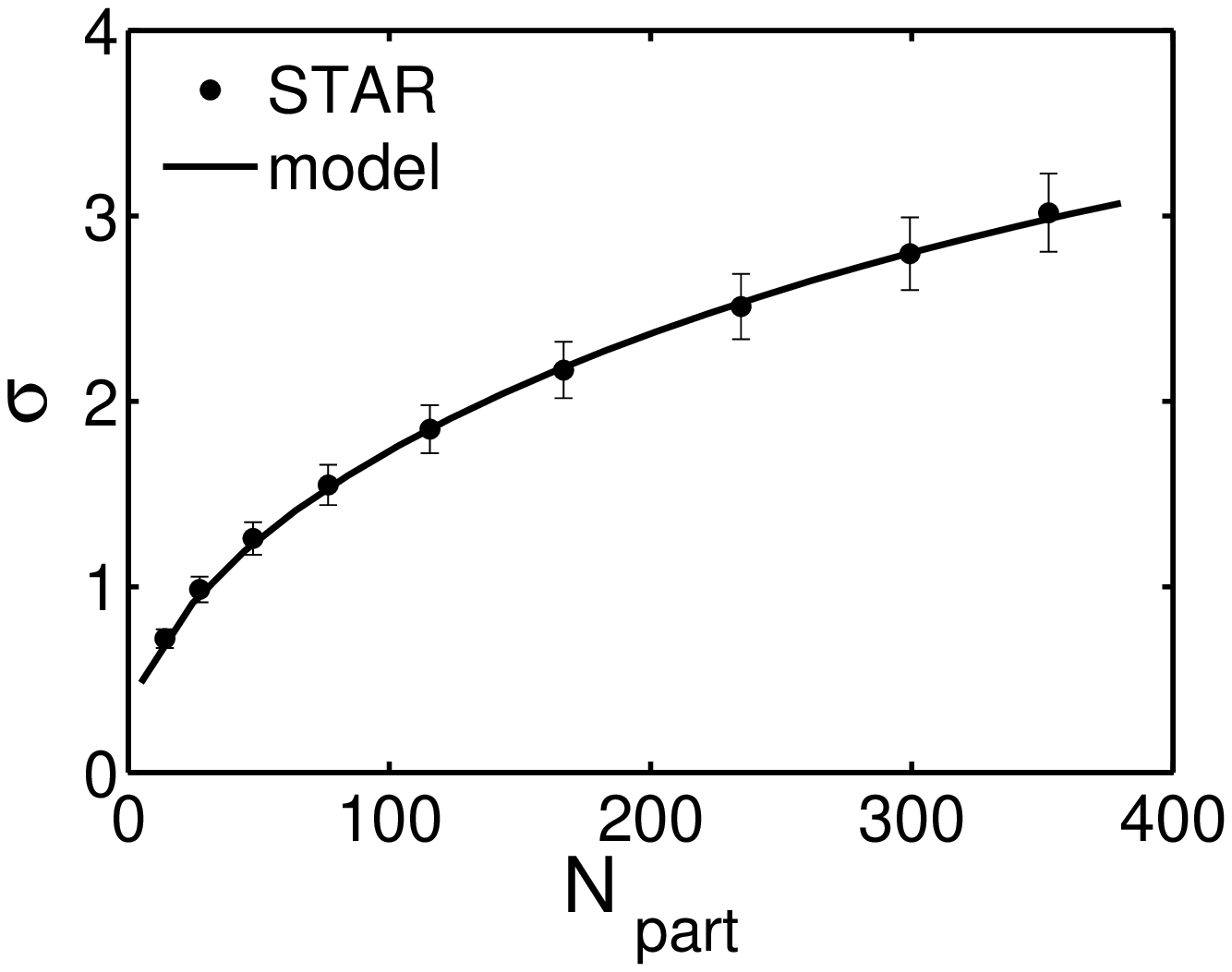}
\caption{The variance for the  net-proton  distributions from multiple emission sources.
The points are from Ref. \cite{STAR}, and the curve is from our model calculation.}
\label{fig4}
\end{figure}

\begin{figure}[tbph]
\includegraphics[width=0.45\textwidth]{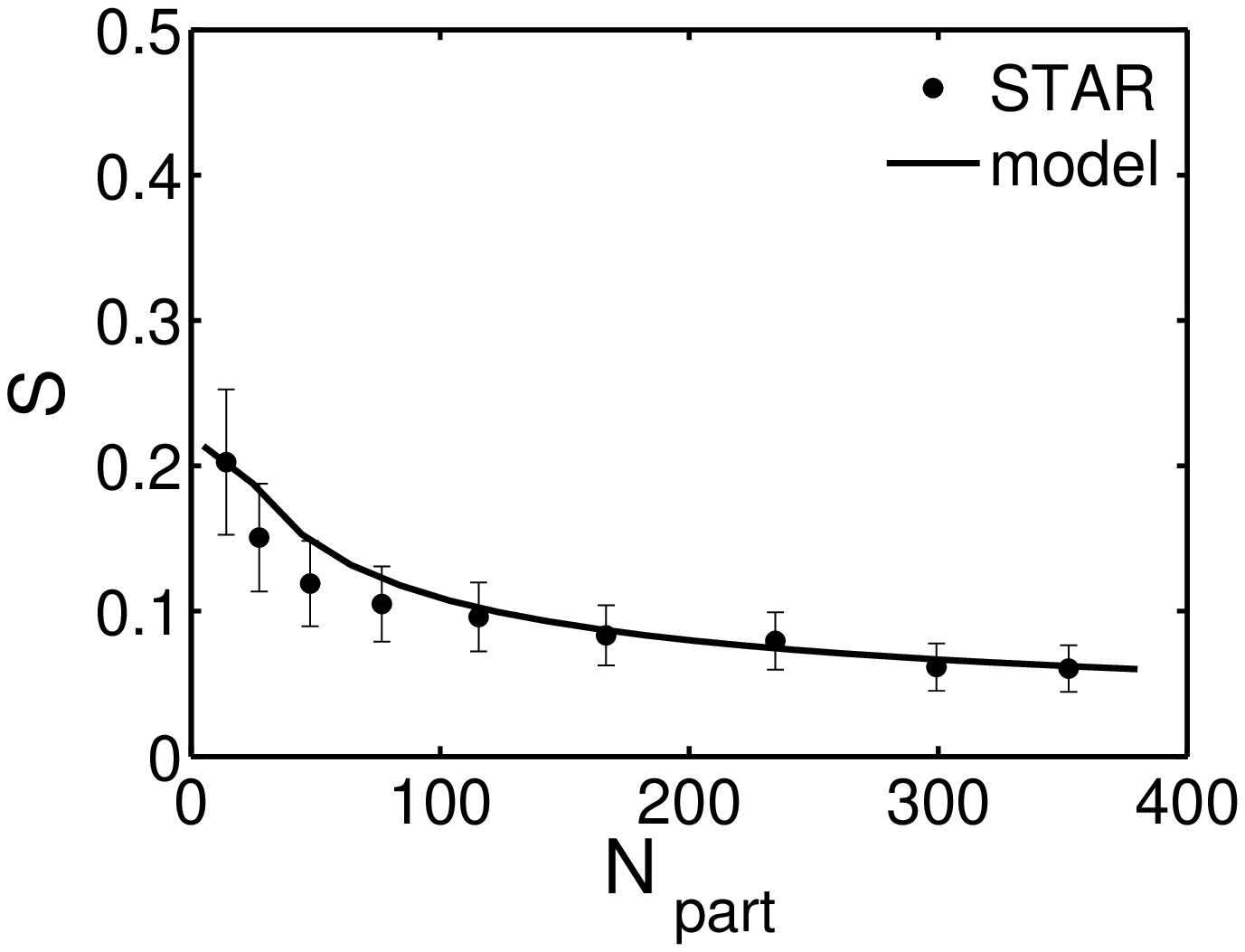}
\caption{The skewness for the  net-proton  distributions from multiple emission sources.
The points are from Ref. \cite{STAR}, and the curve is from our model calculation.}
\label{fig5}
\end{figure}

\begin{figure}[tbph]
\includegraphics[width=0.45\textwidth]{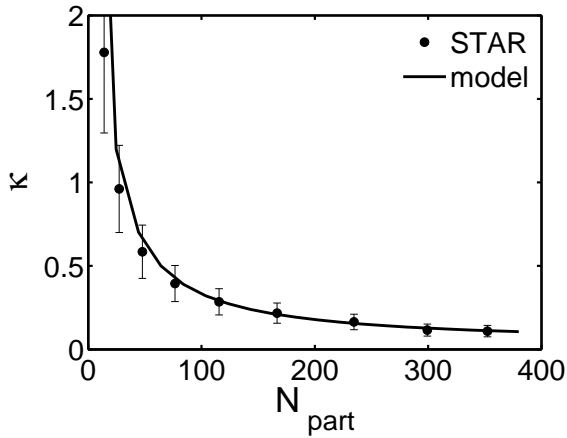}
\caption{The kurtosis for the  net-proton  distributions from multiple emission sources.
The points are from Ref. \cite{STAR}, and the curve is from our model calculation.}
\label{fig6}
\end{figure}

To make predictions for the net-proton distributions at other centralities, one can parameterize the values of parameters $\mu$ and $B\rho$
tabulated in TABLE \ref{tab1} by polynomials of the number of participants $N_p$ as
\begin{eqnarray}
\mu&=& 0.171+0.0495N_p-2.5\times 10^{-5}N_p^2,\\
B\rho&=& (-4.63+5.99N_p-3.7\times 10^{-3})/1000 .
\end{eqnarray}
From this parameterization, one can calculate the moments for the distributions easily.
 The obtained moments are shown, as functions of the number of
participants $N_{\rm part}$, in Figs. \ref{fig3}-\ref{fig6}. The corresponding experimental data from
\cite{STAR} are shown in the figures for comparison. The calculated mean, variance, skewness and kurtosis
are well in agreement with
the data. The good agreement shows that the basic merits for the baryon production mechanism have been
exhibited in our model consideration.

\begin{figure}[tbph]
\includegraphics[width=0.45\textwidth]{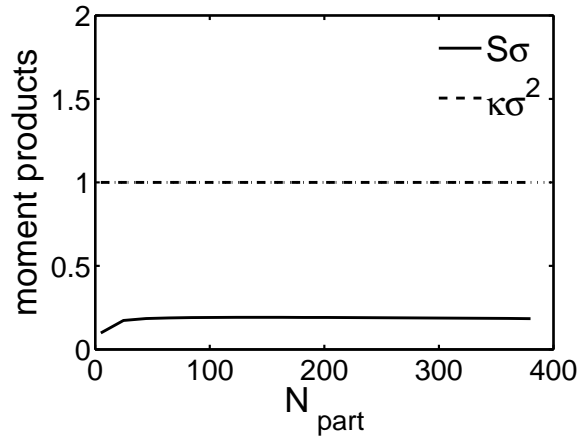}
\caption{The moment products, $S\sigma$ and $\kappa\sigma^2$, for the  net-proton distributions
from multiple emission sources as functions of $N_{\rm part}$. The dotted line is for $\kappa\sigma^2=1$
expected from the hadron resonance gas model.}
\label{prod}
\end{figure}

In Ref. \cite{THEO4}, the observed moments
are related to the ones from one emission source by using the central limit theorem.
This expectation is deduced from the independent emission of particles from each source.
In fact, if the baryons are produced from $N_S$ identical sources, one can get relations of the moments for
the measured distributions and those for the emission sources as
\begin{eqnarray*}
M&=&M_iN_S,\\
\sigma&=&\sigma_i\sqrt{N_S}, \\
S&=&S_i/\sqrt{N_S}, \\
\kappa&=&\kappa_i/N_S,
\end{eqnarray*}
\noindent where quantities with subscript $i$ are for moments from one emission source.
From the above expressions on centrality dependence,
one can expect constant $S\sigma$ and $\kappa\sigma^2$ for all $N_{\rm part}$. In our fitting with
a single emission source, both $\mu$ changes strongly with
centrality. As a result of the centrality dependence of $B\rho$ and $\mu$, the centrality
dependence of the moments are well reproduced, as can be seen from Figs. \ref{fig3}-\ref{fig6}.
One can also calculate moment products $S\sigma$ and $\kappa\sigma^2$. The centrality dependence of the
 products are shown in Fig. \ref{prod}. In the $N_{\rm part}$ range shown, $S\sigma$ increases a few percent, while $\kappa\sigma^2$ is almost exactly 1, as  expected from the hadron resonance gas model \cite{HRG}.

\begin{figure}[tbph]
\includegraphics[width=0.45\textwidth]{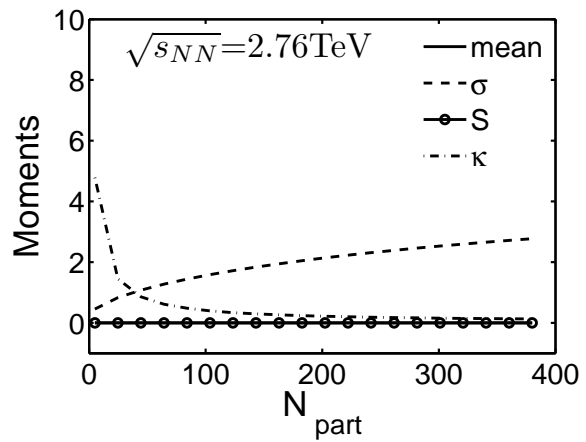}
\caption{The moments for the  net-proton  distribution from multi-emission sources at $\sqrt{s_{NN}}=2.76$
TeV.}
\label{fig7}
\end{figure}

With the good agreement with STAR data at RHIC energy at hand, one can go one step further to predict
the higher order moments of the net-proton distributions at LHC energies. For specific, let the
center-of-mass energy of the colliding nucleon-nucleon pair be 2.76TeV. From fitting data in
\cite{STAR} one can find the center-of-mass  energy dependence of the parameter $B\rho$ and
extrapolate to
 LHC/ALICE energy. In this way, one gets $B\rho\simeq 0$ at $\sqrt{s_{NN}}=2.76$ TeV.
 To get the parameter $\mu$ at LHC/ALICE, one can write $\mu\propto\exp(m_p/T)$, with
  $T$ being the effective emission temperature of the sources and $m_p$ the mass
  of proton. In \cite{Tem} the effective temperature
  of the medium is given as a function of the center-of-mass energy $\sqrt{s}$ as
  $$T=T_{\rm Lim}/[1+\exp(2.6-\ln(\sqrt{s})/0.45)]$$
  \noindent  with $T_{\rm Lim}=164$ MeV. At LHC, the center of mass energy is much higher than at RHIC, so
  the value of parameter $\mu$ is much smaller than obtained in the above.
The smaller value of $\mu$ will allow more baryon
pairs to be produced from a source. The predicted moments are shown in Fig. \ref{fig7}.
Now the mean value of the distribution is zero. Because of zero mean value,
the net-proton distribution at LHC/ALICE is symmetric about 0 and one gets zero value for the skewness.
Since the kurtosis is zero for Gaussian
distributions, extremely small $\kappa$ value at large $N_{\rm part}$ at LHC energies means that the
net-proton distributions at LHC energies can be well parameterized by Gaussian, but not for small
$N_{\rm part}$.

\section{Conclusion}
The higher order moments of the net-proton distributions in relativistic Au+Au collisions
at $\sqrt{s_{NN}}=200 {\rm GeV}$
are studied from a simple model with effects from initial baryon stopping and final baryon pair emission
taken into account. We have demonstrated that by employing a single  emission source model,
the distributions at different collision centralities can be well reproduced.
Then the higher order moments for the distributions can be calculated without new free parameters.
The calculated moments agree well with the experimental results. The predicted moments for LHC
Pb+Pb collisions need to be verified experimentally.

It should be mentioned that nothing else is assumed in this model except an initial net-proton and a
finite probability for producing baryon pairs from sources. Therefore, our model has nothing to do
with  thermal equilibrium and/or critical fluctuations. Because our model consideration is
based on normal physics
effects, our results can be used as a baseline for detecting novel physics in the processes.

\acknowledgments{This work was supported in part by the National Natural Science Foundation of
China under Grant No. 11075061 and by the Programme of Introducing Talents of Discipline to
Universities under No. B08033. The authors thank Dr. X.F. Luo for sending us the experimental data.
We are grateful to N. Xu and X.F. Luo for valuable discussions. }


\begin{thebibliography}{99}
\bibitem{YA} Y. Aoki et al., Nature {\bf 443}, 675 (2006); M. Cheng et al.,
Phys. Rev. {\bf D 74}, 054507 (2006).

\bibitem{JB} J. Berges, K. Rajagopal, Nucl. Phys. {\bf B 538}, 215 (1999).

\bibitem{MAS} M. Stephanov, K. Rajagopal and E. Shuryak, Phys. Rev. {\bf D 60}, 114028 (1999).

\bibitem{FLUC} M.A. Stephanov,K. Rajagopal and E.V. Shuryak, Phys. Rev. Lett. {\bf 81}, 4816 (1998);
S. Jeon, V. Koch, Phys. Rev. Lett. {\bf 85}, 2076 (2000); M. Asakawa, U.W. Heinz and B. M\"uller, Nucl. Phys. {\bf A 698}, 519 (2002); V. Koch, J. Phys. {\bf G 35}, 104030 (2008).

\bibitem{SE} S. Ejiri, F. Karsch and K. Redlich, Phys. Lett. {\bf B 633}, 275 (2006).

\bibitem{STAR} M.M. Aggarwal et al., (STAR Collaboration), Phys. Rev. Lett. {\bf 105}, 022302 (2010).

\bibitem{THEO1}  S. Gupta et al., Science {\bf 332}, 1525 (2011); X.F. Luo, B. Mohanty,
H.G. Ritter and N. Xu, arXiv:1105.5049.

\bibitem{THEO2}F. Karsh and K. Redlich, Phys. Lett. {\bf B 693}, 136 (2011); M. A. Stephanov,
arXiv:1104.1627v1;  B. Friman et al., arXiv:1103.3511.

\bibitem{THEO3} Y. Zhou et al., Phys. Rev. {\bf C 82}, 014905 (2010); K. Xiao et al.,
Chin. Phys. {\bf C 35}, 467 (2011).

\bibitem{THEO4} X.F. Luo, B. Mohanty, H.G. Ritter and N. Xu, arXiv:1001.2847.

\bibitem{NP} Y. Hatta and M.A. Stephanov, Phys. Rev. Lett. {\bf 91}, 102003 (2003).

\bibitem{MAS2} M.A. Stephanov, Phys. Rev. Lett. {\bf 102}, 032301 (2009).

\bibitem{LUO} X.F. Luo et al., J. Phys.{\bf G 37}, 094061 (2010).

\bibitem{FK} F. Karch and K. Redlich, Phys. Lett. {\bf B 695}, 136 (2011).

\bibitem{HRG} P. Braun-Munzinger, K. Redlich and J. Stachel, nucl-th/0304013; A. Andronic,
P. Braun-Munzinger and J. Stachel, Nucl. Phys. {\bf A 772}, 167 (2006).

\bibitem{chen} L.Z. Chen et al., arXiv:1011.0712.

\bibitem{begun} V.V. Begun et al., Phys. Rev. {\bf C 70}, 034901 (2004).
\bibitem{Tem} A. Andronic et al., arXiv:1106.6321v1.

\end{thebibliography}
\end{document}